\documentclass[preprint]{aastex}
\usepackage{graphics}

\newcommand {\boom}{{\sc Boom\-erang} }

\newcommand {\bmn}{B98} 
\newcommand {\bm}{B98 }
\newcommand {\bk}{BOOM03 } 
\newcommand {\bkn}{BOOM03}

\begin{document}

\title{Measuring CMB Polarization with {\bf {\sc Boom\-erang}}}

\author{
T.~Montroy\altaffilmark{1,10},  
P.A.R.~Ade\altaffilmark{3}, 
A.~Balbi\altaffilmark{18},
J.J.~Bock\altaffilmark{4,16},
J.R.~Bond\altaffilmark{5},
J.~Borrill\altaffilmark{6,19},
A.~Boscaleri\altaffilmark{7},
P.~Cabella\altaffilmark{18},
C.R.~Contaldi\altaffilmark{5},
B.P.~Crill\altaffilmark{8},
P.~de~Bernardis\altaffilmark{2},
G.~De~Gasperis\altaffilmark{18},
A.~de~Oliveira-Costa\altaffilmark{14},
G.~De~Troia\altaffilmark{2},
G.~di~Stefano\altaffilmark{13},
K.~Ganga\altaffilmark{9},
E.~Hivon\altaffilmark{9},
V.V.~Hristov\altaffilmark{16},
A.~Iacoangeli\altaffilmark{2},
A.H.~Jaffe\altaffilmark{17},
T.S.~Kisner\altaffilmark{1,10},
W.C.~Jones\altaffilmark{16},
A.E.~Lange\altaffilmark{16},
S.~Masi\altaffilmark{2},
P.D.~Mauskopf\altaffilmark{3},
C.~MacTavish\altaffilmark{15},
A.~Melchiorri\altaffilmark{2},
F.~Nati\altaffilmark{2},
P.~Natoli\altaffilmark{18},
C.B.~Netterfield\altaffilmark{15},
E.~Pascale\altaffilmark{15},
F.~Piacentini\altaffilmark{2},
D.~Pogosyan\altaffilmark{11},
G.~Polenta\altaffilmark{2},
S.~Prunet\altaffilmark{12},
S.~Ricciardi\altaffilmark{2},
G.~Romeo\altaffilmark{13},
J.E.~Ruhl\altaffilmark{1},
E.~Torbet\altaffilmark{10},
M.~Tegmark\altaffilmark{14}, and
N.~Vittorio\altaffilmark{18}.
}

\affil{
$^1$ Physics Department, Case Western Reserve University,
		Cleveland, OH, USA\\
$^2$ Dipartimento di Fisica, Universit\'a di Roma La
Sapienza, Roma, Italy \\
$^{3}$ Dept. of Physics and Astronomy, Cardiff University, 
		Cardiff CF24 3YB, Wales, UK \\
$^4$ Jet Propulsion Laboratory, Pasadena, CA, USA\\
$^5$ Canadian Institute for Theoretical Astrophysics, 
		University of Toronto, Toronto, Ontario, Canada\\
$^6$ National Energy Research Scientific Computing Center, 
		LBNL, Berkeley, CA, USA\\
$^7$ IFAC-CNR, Firenze, Italy\\
$^8$ CSU Dominguez Hills, Carson, CA, USA\\
$^9$ IPAC, California Institute of Technology, Pasadena, CA, USA\\
$^{10}$ Dept. of Physics, University of California, Santa Barbara, CA, USA\\
$^{11}$ Physics Dept., University of Alberta, Edmonton, Alberta, Canada\\
$^{12}$ Institut d'Astrophysique, Paris, France\\
$^{13}$ Istituto Nazionale di Geofisica, Roma,~Italy\\
$^{14}$ Physics Department, University of Pennsylvania, Philadelphia, PA, USA\\
$^{15}$ Physics Department, University of Toronto, Toronto, Ontario, Canada\\
$^{16}$ Observational Cosmology, California Institute of
Technology, Pasadena, CA, USA\\
$^{17}$ Astrophysics Group, Imperial College, London, UK\\
$^{18}$ Dipartimento di Fisica, Universit\'a di Roma Tor
Vergata, Roma, Italy\\
$^{19}$ Center for Particle Astrophysics, University of California,
Berkeley, CA, USA\\
}

\begin{abstract}
\boom is a balloon-borne telescope designed for long duration (LDB)
flights around Antarctica. 
The second LDB Flight of \boom took place in
January 2003. The primary goal of this flight was to measure the
polarization of the CMB. The receiver uses polarization sensitive
bolometers at 145 GHz. Polarizing grids provide polarization
sensitivity at 245 and 345 GHz.
We describe the \boom telescope noting changes made
for 2003 LDB flight, and discuss some of the issues involved in
the measurement of polarization with bolometers. Lastly, we report on
the 2003 flight and provide an estimate of the expected results.
\end{abstract}


\section{Introduction}

The 1998 flight of \boom (\bmn) provided a measurement of the angular power
spectrum of CMB temperature
anisotropies from $\ell= 25$ to
$\ell = 1000$ \citep{ruhl_etal}. \boom made its second long duration balloon (LDB) flight (\bkn) in January
2003 with a receiver configured to
simultaneously measure CMB temperature and polarization anisotropies. The new receiver
used pairs of polarization sensitive bolometers (PSB's) at 145
GHz. 
At 245 GHz and 345 GHz, spider web bolometers are used with polarization sensitivity provided by
polarizing grids mounted at the front of the cryogenic feed
horns. The \bm results were partially limited by pointing
reconstruction error ($2.5^{\prime}$ rms). For the \bk flight, we
added a pointed sun sensor and a tracking star camera which should
reduce our pointing reconstruction error to less than $1^{\prime}$ rms.
From the \bk flight, we have 11.7 days of data. An analysis effort is
underway, with the goal of producing measurements of $C_{\ell}^T$, $C_{\ell}^{TE}$ and  $C_{\ell}^{EE}$.

\section{Telescope}

\begin{figure}[!t]
{\par\centering \resizebox*{0.8 \textwidth}{!}{\rotatebox{0}
{\includegraphics[0.5in,0.0in][5in,4.5in]{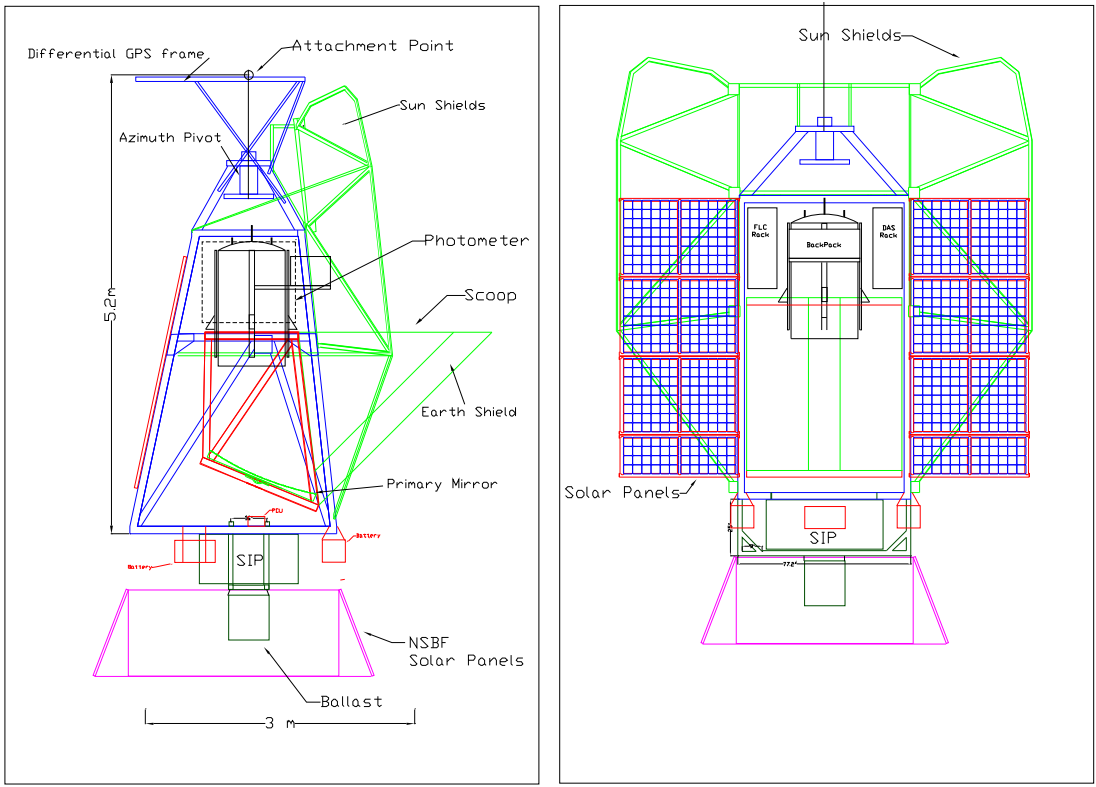}}} \par}
\caption{\small 
The \boom~payload.  The extensive shielding prevents stray-light 
contamination and helps to maintain thermal stability.
\label{fig:tele}}
\end{figure}


The \boom telescope (Figure \ref{fig:tele}) was designed specifically for the harsh conditions
of Antarctic ballooning \citep{crill02}. 
The extensive shielding prevents the contamination by
stray-light, and protects other sensitive components from the Sun.
This is especially important for daytime ballooning over Antarctica,
where thermal management is vital. The back of the telescope is constantly
illuminated by the Sun, reaching temperatures of $55^{\circ}$ C, while the
front is shaded and can cool to less than $-20^{\circ}$ C.

\boom scans in azimuth, with the 
elevation 
kept constant for at least one hour. Sky rotation turns
this one-dimensional azimuthal scan strategy into a cross-linked
pattern on the celestial sphere. 
\bm and \bk both used a differential GPS array, a fixed Sun
sensor and rate gyros for pointing reconstruction. For \bkn, the
pointed sun sensor and tracking star camera should improve
the pointing reconstruction error.

\begin{figure}[!ht]
{\par\centering \resizebox*{0.9 \columnwidth}{!}{\rotatebox{0}{\includegraphics{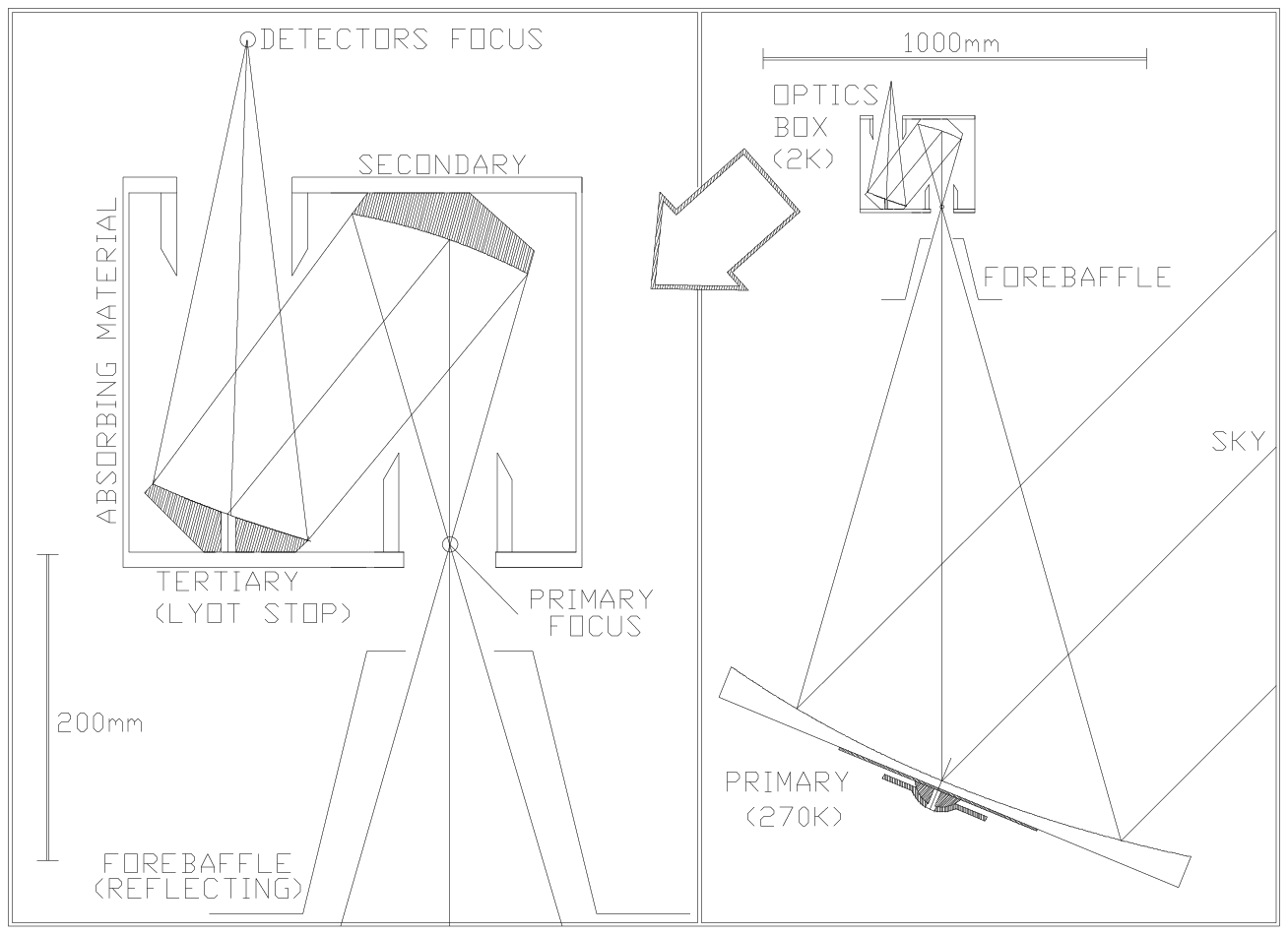}}} \par}
\caption{\small~An overview of the \boom~optics. Radiation from the primary reflects
off the secondary, to the tertiary and then to the focal plane. The
secondary and tertiary correct aberrations induced by the
primary. They are kept at 1.65 K in a box coated with absorbing material. The tertiary acts as the Lyot stop for the system,
controlling the illumination of the primary. 
Spillover off the edge of the tertiary sees a 1.65 K blackbody.
\label{fig:optics}}
\end{figure}

The \boom primary mirror is an off-axis parabola with a diameter of
1.3~m. It is 45$^\circ$ degrees off axis and has a focal length of 1280~mm. 
The primary feeds a pair of cold re-imaging mirrors which are 
kept at 1.65~K inside the cryostat (Figure \ref{fig:optics}). The
tertiary forms an image of the primary and acts as the Lyot stop for our
system. It controls the illumination on the primary, limiting the
effective diameter to about 80 cm. There is a  1 cm hole in the
tertiary behind which sits a calibration lamp. It fires once every
15 minutes allowing us to monitor any calibration drift.

The \boom cryostat can keep the detectors at 0.275 K for 15 days
\citep{cryo2}. Toroidal nitrogen
and helium tanks (with the helium tank inside the nitrogen tank) are 
suspended via 1.6 mm Kevlar ropes. The insert, which contains the cold
mirrors and the focal plane, is bolted to the helium tank. The insert
also contains a single stage $^3$He refrigerator which can keep
the focal plane at 0.275 K for 12-13 days \citep{cryo2}. Since the
$^4$He and LN$_2$ stages have a longer hold time (18 and 16 days
respectively) than the $^3$He stage, we
added capability for an in-flight cycle of the $^3$He refrigerator. 

\section{\bk Receiver}

\begin{figure}[t]
{\par\centering \resizebox*{1.0 \textwidth}{!}{\rotatebox{90}
{\includegraphics[0.5in,2.5in][5in,10in]{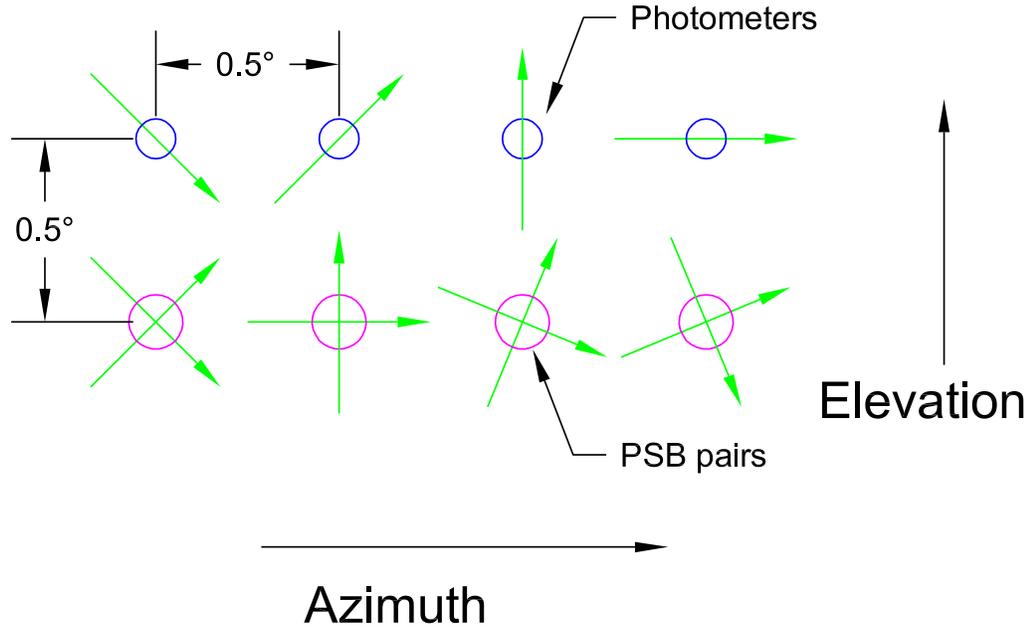}}} \par}
\caption{\small
Focal Plane Schematic. 2-color photometers with band centers at 245
GHz and 345 GHz populate the upper row. Each photometer is only
sensitive to one polarization. The lower row has 4 pairs of PSB's. 
The elements in a PSB pair are sensitive orthogonal polarizations. The circles
representing the pixels show relative beams sizes: $\sim 7^{\prime}$ for both
photometer channels and $9^{\prime}$ for the PSB's. The green arrows through
the circles show the orientation of the polarization sensitivity. The
photometer and PSB rows are separated by $0.5^{\circ}$ in elevation,
while the pixels in a row are separated by $0.5^{\circ}$ in
cross-elevation.
\label{fig:focalplane}}
\end{figure}

The \bk receiver was designed for measurement of CMB
temperature and polarization anisotropies. There are 8 pixels, and each pixel contains
two detectors.
Four of the pixels contain pairs of
polarization sensitive bolometers (PSB's) operating at 145 GHz. The
other four pixels are 2-color photometers operating at 245 and 345
GHz. Both the PSB's and the photometers use corrugated feed horns.
Table \ref{tab:rec} summarizes the properties of the \bk receiver.
Figure \ref{fig:focalplane} shows the layout of the focal plane
projected onto the sky.

\begin{table}[!h]
\begin{center}
\begin{tabular}{|c|c|c|c|c|}
\hline
Freq & Bandwidth & $\#$Channels & Beam FWHM  & Exp. NET$_{CMB}$\\ 
\hline
145 GHz & 46 GHz & 8 & $9.5^{\prime}$  & 160 $\mu K \sqrt{s}$\\
245 GHz & 100 GHz & 4 & $6.5^{\prime}$ & 290 $\mu K \sqrt{s}$\\
345 GHz & 100 GHz & 4 & $7^{\prime}$ & 660 $\mu K \sqrt{s}$ \\
\hline
\end{tabular}
\end{center}
\begin{center}
\caption{\small
Summary of the properties of the \bk receiver. The FWHM is based on a pre-flight
measurement using a tethered thermal source placed 1~km from the
telescope. The $NET_{CMB}$ is calculated using a model of our
detectors based on pre-flight loadcurve data. Preliminary analysis
of flight data provides a similar answer at 145~GHz 
\label{tab:rec}}
\end{center}
\end{table} 

\subsection{Polarization Sensitive Bolometers}

\begin{figure}[!t]
\begin{center}

\resizebox*{1.0 \columnwidth}{!}{\rotatebox{0}{\includegraphics{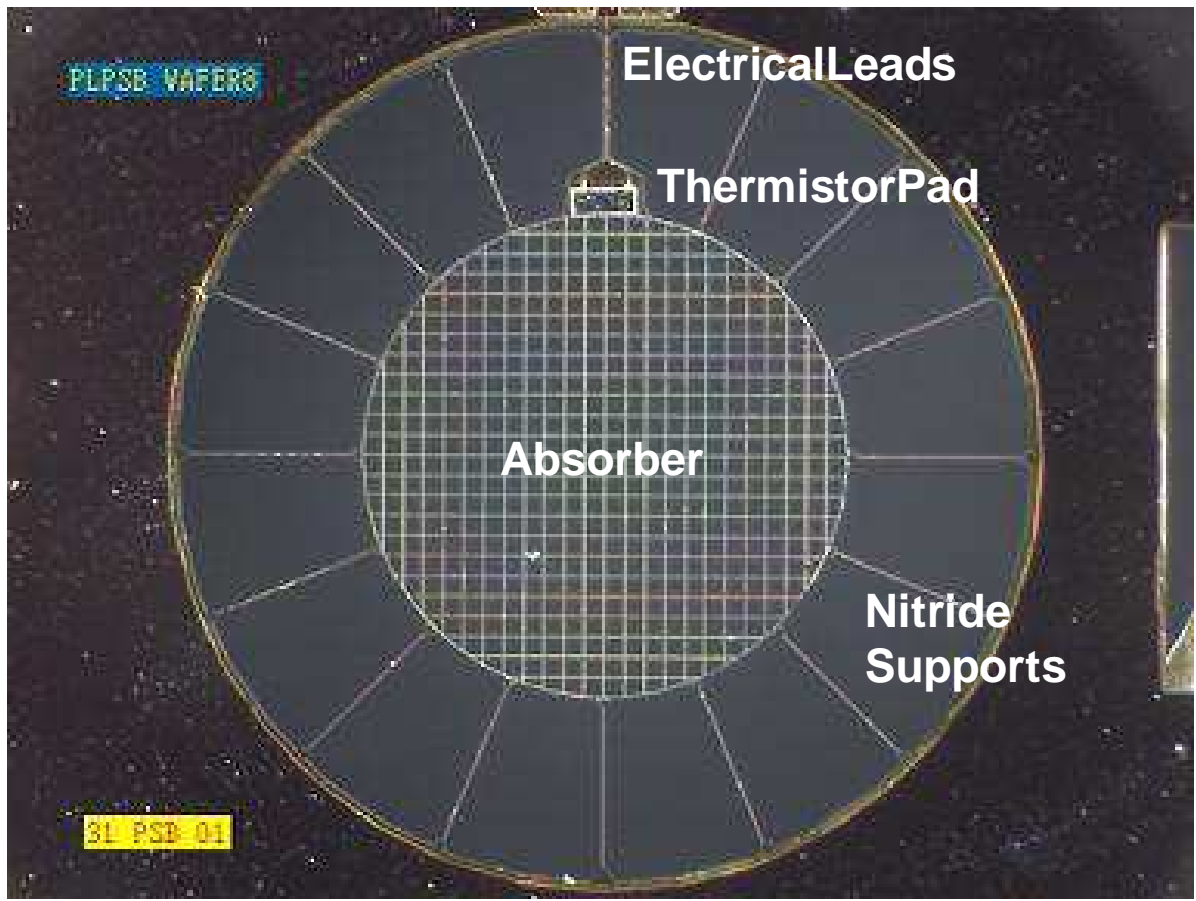}}}

\caption{\small 
A Polarization Sensitive Bolometer. The diameter of the grid is
2.6~mm. The Si$_3$N$_4$ mesh is metallized in the vertical direction
only, making the device sensitive to vertical polarization only. The
horizontal component of the grid is un-metallized and provides
mechanical support for the device. Orthogonally metallized devices are
spaced 
$60 \mu$m apart at the exit aperture of a corrugated feed structure.
\label{fig:psb}}
\end{center}
\end{figure}

The polarization sensitive bolometers are a variation on the original micro-mesh
design used in \bm \citep{psbs}. Instead of a spider web design, the
mesh is a square grid. 
The grid is only metallized in one direction (Figure
\ref{fig:psb}) making it sensitive
to only one component of the incoming electric field.  A pair of these with 
metallized directions oriented $90^{\circ}$ apart are
mounted at the end of a corrugated feed 
structure, separated by $60\mu$m along the axis of propagation. This allows for simultaneous measurement of both electric
field components at the same point on the sky, through the same
filters and feed structure.

\begin{figure}[t]


{\par\centering \resizebox*{0.75 \columnwidth}{!}{\rotatebox{0}
{\includegraphics[1.25in,3.25in][6.5in,7.5in]{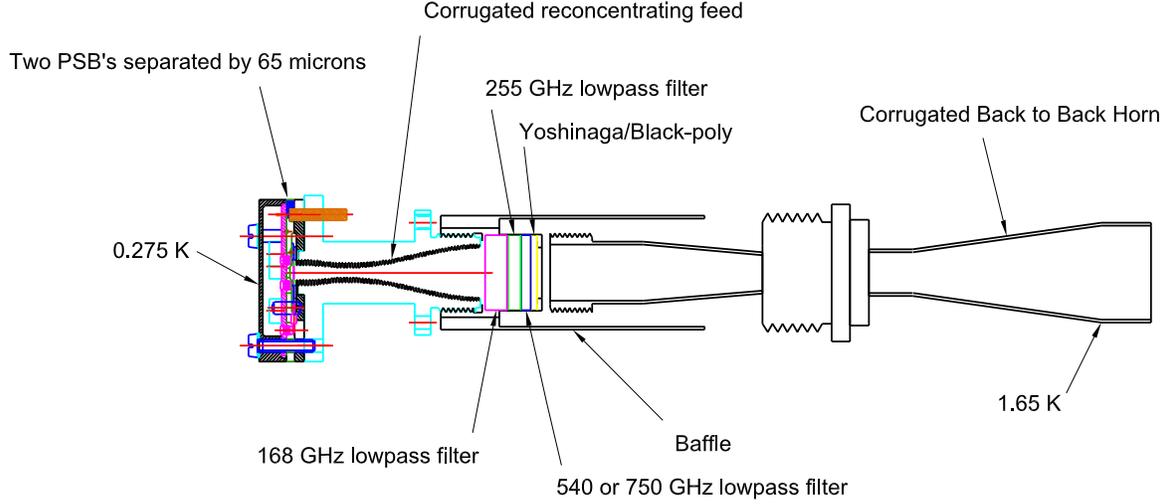}}} \par}
\caption{\small 
The PSB feed structure. Radiation travels through the back-to-back
feed and passes through the filter stack as it enters the
reconcentrating feed. The PSB pair sits at the exit aperture of the
reconcentrating feed. Both the back-to-back feed and the reconcentrating
feed are designed to preserve polarization.
\label{fig:psbfeed}}

\end{figure}

Figure \ref{fig:psbfeed} shows the feed structure for a PSB pair. Light
enters into a corrugated back-to-back feed horn, travels out of the
back feed horn, across
the thermal break and into the filter stack which is mounted on the
front of the corrugated reconcentrating feed. The filter stack
consists of metal mesh low pass filters which which define the upper edge of the band to
be roughly $170$ GHz and a
Yoshinaga/Black-poly filter which prevents high frequency leaks.
The lower edge of the band is defined by the waveguide cut-off of
the back-to-back horn to be $122$~GHz. Once the radiation passes through the
filter stack, it  enters the reconcentrating feed which 
couples the radiation to the pair of PSB's sitting at its exit aperture. The entire feed
structure is designed so that polarization is preserved as the
radiation travels from the entrance aperture to the bolometers \citep{psbs}.  

\section{Photometers}
\begin{figure}[!t]

{\par\centering \resizebox*{0.6  \columnwidth}{!}{\rotatebox{0}
{\includegraphics[1.25in,3.25in][6.5in,7.5in]{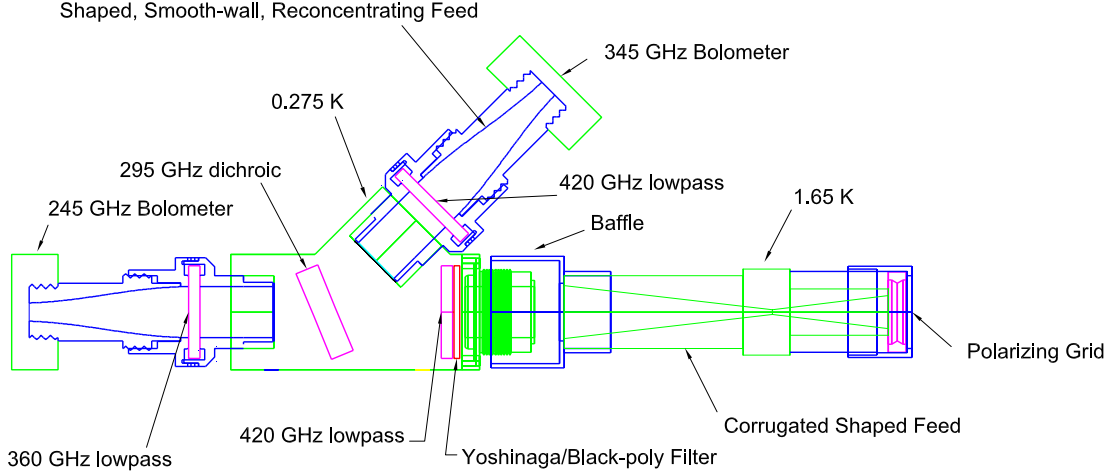}}} \par}

\caption{\small 
The photometer feed structure. Radiation passes through a polarizing
grid mounted in front of the back-to-back feed horn. It exits the feed
horn and enters the photometer body passing through a pair of low pass
filters. A dichroic transmits frequencies below 295 GHz to the 245 GHz
bolometer and reflects high frequency radiation to the 345 GHz bolometer.
\label{fig:photometer}}
\end{figure}

The  2-color photometer (Figure \ref{fig:photometer}) design has
evolved from the 3-color photometer of \bmn. The photometers operate at
245 and 345~GHz using conventional spider web bolometers. It is fed by a
back-to-back corrugated feed which was designed to be
single-moded from 180~GHz to nearly 400~GHz. The photometers are made
polarization sensitive by placing
a polarizing grid in front of feed horn entrance aperture. 

Incoming radiation passes through the polarizing grid and into the
back-to-back 
feed horn. It exits the feed horn and enters the
photometer body passing through a metal mesh low pass and a
Yoshinaga/Black-poly filter. Radiation below 295 GHz is transmitted by the
dichroic to the 245 GHz bolometer, and the high frequency radiation is
reflected to the 345 GHz bolometer.

\section{Measuring Polarization}

The Stokes parameters completely describe the polarization state of an
electric field. A general polarized electromagnetic wave with angular
frequency $\omega = 2\pi\nu$ can
be described by:
\begin{equation}
\vec{E} = E_{x}(t) sin(\omega t -\delta_{x}(t)) \hat{x} + E_{y}(t)sin(\omega t -\delta_{y}(t)) \hat{y}.
\label{eq:efield}
\end{equation}
The Stokes parameters can be written as
\begin{eqnarray}
I &=& \left<E_{y}^{2} + E_{x}^{2}\right>,\\
Q &=& \left<E_{x}^{2} - E_{y}^{2}\right>,\\
U &=& \left<2 E_{y} E_{x} \cos(\delta_{y} - \delta_{x})\right>,\\
V &=& \left<2 E_{y} E_{x} \sin(\delta_{y} - \delta_{x})\right>,\\
\tau &=&  \frac{1}{2}\tan^{-1}(\frac{U}{Q}),
\end{eqnarray}
where the averaging is done over times scales longer than the period of
the wave. The total intensity of the radiation is described by $I$. Parameters $Q$
and $U$ describe the linear polarization, while $V$ quantifies the degree
of circular polarization ($V = 0$ when the radiation is linearly polarized). 
$\tau$ is the polarization angle.

Ideally, the signal from one of our detectors
is comprised of the total power received in one polarization. For example, the
signal received by a detector sensitive to the x-component of the electric
field would be proportional to $E_x^2$. In order to measure $Q$ and
$U$, we must combine data from different detectors or multiple
rotation angles for a single detector. This issue is
complicated by the fact that each of our detectors has a finite
cross-polarization ($1\% - 10\%$) meaning that our x-polarized
detector is contaminated with a small amount of $E_y^2$.

A single detector with no cross-polarization 
measures a combination $I$, $Q$ and $U$ in a pixel $p$. The signal is
\begin{eqnarray}
V_{i}^p &=& \gamma_i (I^p + Q^p \cos{(2\alpha)} + U^p \sin{(2\alpha})),\\ 
      &=& \gamma_i T_i^p
\end{eqnarray}
where $V_{i}$ is the detector voltage, $\alpha$ is the polarization orientation of the detector,
$\gamma_i$ is the calibration factor. $I$, $Q$ and
$U$ are expressed as deviations from $T_{CMB}$ and $T_i^p$ is the total signal
seen in one polarization. The
calibration factor (with units $V/K_{CMB}$) is defined with respect to an unpolarized source. This is
consistent with defining CMB temperature and polarization signals as 
\begin{eqnarray} 
T &=& \frac{T_x + T_y}{2},\\
Q &=& \frac{T_x - T_y}{2}.
\end{eqnarray}

The polarization efficiency can be described as the
ratio of the signal in the desired polarization to total signal, for example
\begin{equation}
\rho_x = \frac{2\gamma_x E_{x}^2}{V_{x}},
\end{equation}
where $E_{x}^2$ is the x-component of the incoming radiation and $V_{x}$ is
the signal 
measured by a detectors with polarization orientation in the $\hat{x}$
direction. If $\rho = 1$, there is no cross-polarization. 
If we rotate a polarized source in front of our detector, then the
detector sees a signal
\begin{equation}
S  = \beta (1-\rho \sin^2(\theta_{det}-\theta_{source})),
\end{equation}
where $S  = \beta$ when the source and detector have aligned
polarizations and $S = \beta(1-\rho)$ when the source and
detector are $90^{\circ}$ apart.

To understand the effect of polarization efficiency, it is
easiest to consider a PSB pair
with polarizations oriented in the $\hat{x}$ and $\hat{y}$
directions. For the general case, we can apply the rotation rules for the
Stokes parameters.
Including the effects of cross polarization, we have
\begin{eqnarray}
V_{x} &=& 2 \gamma_x (E_x^2 + (1-\rho_x) E_y^2), \\
V_{y} &=& 2 \gamma_y (E_y^2 + (1-\rho_y) E_x^2),
\end{eqnarray}
which can be re-written as
\begin{eqnarray}
V_{x} &=& 2 \gamma_x ((1-\frac{\rho_x}{2}) I + \frac{\rho_x}{2} Q), 
\label{eq:vx}
\\
V_{y} &=& 2 \gamma_y ((1-\frac{\rho_y}{2}) I - \frac{\rho_y}{2} Q) .
\label{eq:vy}
\end{eqnarray}
If we let $\rho_x = \rho_y = \rho$, then we get a rather simple solution
\begin{eqnarray}
I &=& \frac{1}{2} \frac{\frac{V_x}{\gamma_x} + \frac{V_y}{\gamma_y}}{2-\rho},\\
Q &=& \frac{1}{2} \frac{\frac{V_x}{\gamma_x} - \frac{V_y}{\gamma_y}}{\rho},
\end{eqnarray}
illustrating that the cross-polarization causes a loss of efficiency
for $Q$ measurement. Since the polarization signal is approximately $10\%$ of the
temperature anisotropy signal, the above equations illustrate the importance of
high-precision measurements of the gains and polarization efficiencies
for each channel.

\section{\bk Flight}

The second long duration flight of \boom was launched on January 6,
2003 from Williams Field at McMurdo Station, Antarctica. The flight
lasted for 15 days 
before it was terminated on January
21. During that time period, we were able to get 11.7 days of good
data. We lost 20 hours because the $^3$He refrigerator ran out at the end
of day 11, and we had to recycle it. Also, the payload was losing
altitude for most of the flight. At the end of day 13, we had to shutdown the
telescope because wind shear at 70,000~ft made attitude control
difficult. 

Even with the loss of altitude from 120,000 ft to 70,000
ft, the 145~GHz channels were not severely effected. At
75,000~ft, the 145~GHz channels had a responsivity which
was approximately $15\%$
less than the peak responsivity, while the 245 and 345~GHz channels had
responsivities which were $30\%$ less than their peak value. We did not dip below 95,000~ft until the
end of day 11. At 95,000 ft, the 145~GHz channels had only a $5\%$ responsivity
loss, while the 245 and 345~GHz channels had a $10\%$ loss.

\begin{figure}[!t]
{\par\centering \resizebox*{0.9 \columnwidth}{!}{\rotatebox{90}
{\includegraphics{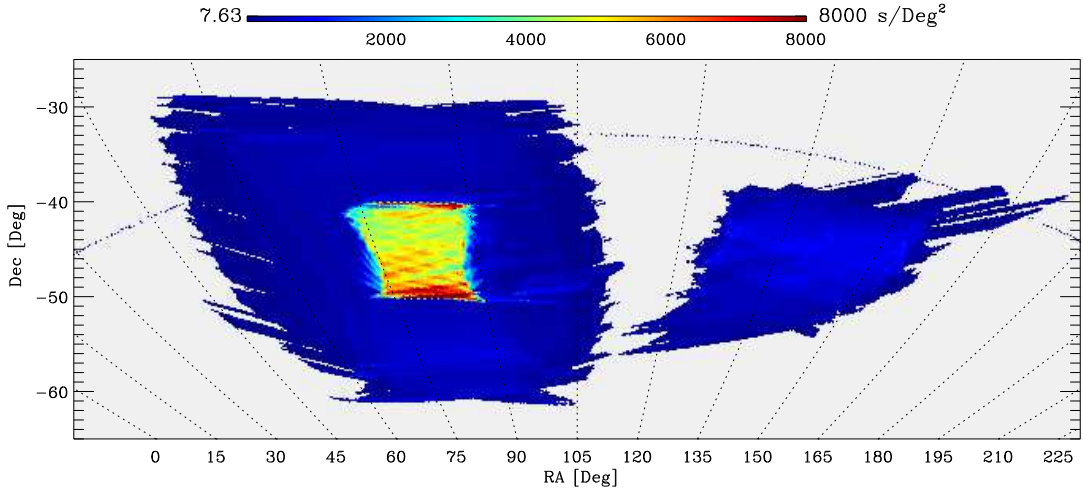}}} \par}

\caption{\small
Sky coverage for one element of one PSB pair. 
\label{fig:b2k2_coverage}}

\end{figure}

The scan strategy was designed to balance the effects of sample
variance and noise, optimizing our sensitivity to
$C_{\ell}^T$, $C_{\ell}^{TE}$ and  $C_{\ell}^{EE}$. Our entire CMB
region covers an area of 1284 deg$^2$, slightly larger than the region
used for \bm in \cite{ruhl_etal}. This region was
split into a shallow region of size 1161 deg$^2$ and a deep region of
size 123 deg$^2$. We also spent time observing near the galactic
plane in an attempt to characterize foreground polarization. Figure
\ref{fig:b2k2_coverage} shows a plot of the sky coverage for one 145~GHz
detector and Table \ref{tab:coverage} shows how our observation time was divided. 

\begin{table*}[!h]
\begin{center}
\begin{tabular}{|c|c|c|c|}
\hline
Region & 7' pixels & Area (deg$^2$) & Average Time (s) \\
CMB Deep  & 9397 & 123 & 46 \\
CMB Shallow & 88547 & 1161 & 3.35\\
CMB Total & 97944 & 1284 & 7.45 \\
Galaxy & 29944 & 393 & 4.67 \\
\hline
\end{tabular}
\end{center}
\caption{\small
A list of the \bk primary scan regions, including the average amount of
observation time per pixel per detector. The CMB Total region is the union of the
CMB Shallow and CMB Deep regions; the average time per pixel is a bit skewed
since the deep region pixels dominate the average.
\label{tab:coverage}}
\end{table*} 

\section{Projected Results}

\begin{figure}[!t]
\begin{center}
\resizebox{5in}{!}{
  \includegraphics[0.5in,2.0in][8in,10in]{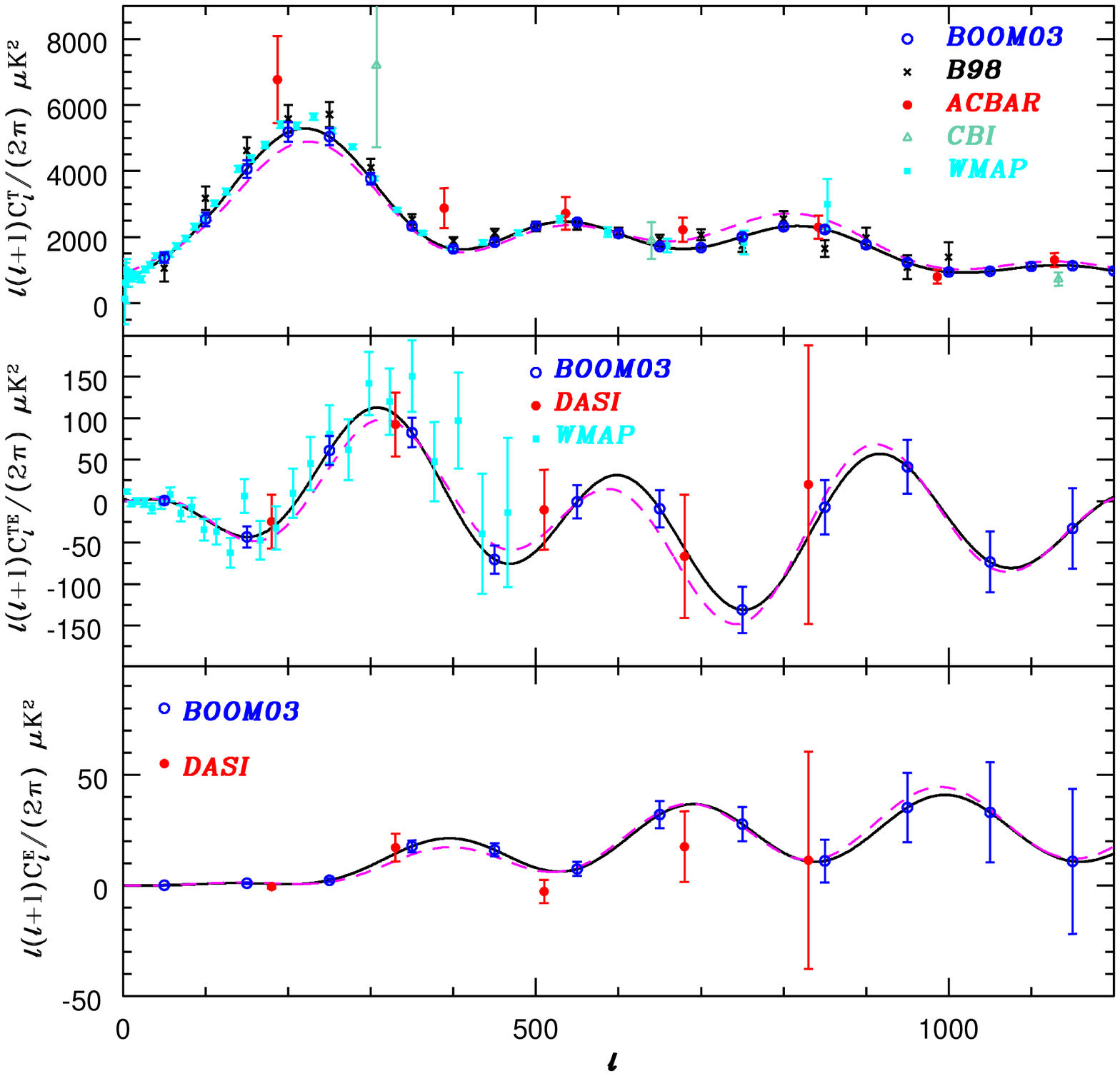}
}
\end{center}
\caption{\small 
Forecasted results for \bk at 145 GHz compared with recent
results from \bm \citep{ruhl_etal}, ACBAR \citep{acbar}, WMAP \citep{wmap_hinshaw}, CBI
\citep{cbi} and DASI \citep{dasi_pol_results}. The top panel shows $C_{\ell}^T$, the middle shows
$C_{\ell}^{TE}$, and the bottom $C_{\ell}^{E}$. The models are 
best fits to the pre-WMAP data \citep{goldstein} for a $\Lambda CDM$ model (black line)
and a $\Lambda = 0$ model (dashed magenta line).  This is a statistical
calculation assuming that all 8 145~GHz channels have an NET$_{CMB} = 160 \mu K \sqrt{s}$. No
allotment is made for calibration uncertainty, beam uncertainty,
pointing error, systematics or the
fact that our long time constants at 145 GHz will decrease signal-to-noise at high-$\ell$.
\label{fig:b2k2_forecast}}
\end{figure}

With our flight data in hand, we can project our sensitivity to
$C_{\ell}^T$, $C_{\ell}^{TE}$, and  $C_{\ell}^{EE}$. Figure
\ref{fig:b2k2_forecast} shows what we can expect statistically with eight detectors at
145 GHz. 
These error bars are estimated
using the formulas presented in \cite{zss97}. The combined NET assumes
that all channels have an  NET$_{CMB} = 160 \mu K \sqrt{s}$. No
allotment is made for calibration uncertainty, beam uncertainty,
pointing error, systematics or the
fact that long time constants at 145 GHz will decrease signal-to-noise at high-$\ell$.

\section{Acknowledgments}
The \boom project has been supported by NASA, NSF-OPP and NERSC in the
U.S., by PNRA, Universit\'a ``La Sapienza'' and ASI in Italy, by
PPARC in the UK, and by CIAR and NSERC in Canada. T.M. acknowledges
support from a NASA GSRP fellowship. The authors would like to thank
the National Scientific Balloon Facility (NSBF) and the U.S. Antarctic Program
for excellent field and flight support.

\bibliographystyle{apj}

\end{document}